\documentclass[10pt]{article}
\usepackage{mathrsfs}
\usepackage{multicol}
\usepackage{amsmath}
\usepackage{graphicx}
\usepackage{float}
\usepackage{flafter}
\usepackage{amsmath}
\usepackage{latexsym}
\usepackage{cite}
\usepackage{bm}
\usepackage{indentfirst}
\usepackage{balance}

\newcommand{\bee}{\begin{equation}}
\newcommand{\ee}{\end{equation}}
\newcommand{\beea}{\begin{eqnarray}}
\newcommand{\eea}{\end{eqnarray}}

\begin{document}

\title{Electron spin-orbit interaction in helically coiled carbon nanotube}
\author{{Ning Ma}$^{1,2\dag}$, {Daqing Liu}$^{2}$, {Vei Wang}$^{2,3}$, {Erhu Zhang}$^{2}$, {%
Shengli Zhang}$^{{2*}}$ \\
%EndAName
$^{1}$Department of Physics, MOE Key Laboratory of Advanced Transducers \\
and Intelligent Control System, Taiyuan University of Technology, Taiyuan
030024, China\\
$^{2}$Department of Applied Physics, MOE Key Laboratory for Nonequilibrium
Synthesis \\
and Modulation of Condensed Matter, Xi'an Jiaotong University, Xi'an 710049,
China}
\date{\today}
\maketitle

\begin{abstract}
Recent theoretical and experimental works on carbon nanotubes (CNTs) have
revealed that spin-orbit interaction (SOI) is more robust than it was
thought. Motivated by this, we investigate the SOI in helically coiled CNTs.
Calculations are performed within the tight-binding model with the inclusion
of a four-orbital basis set; thereby the full symmetry of the helical
lattice and the hybridization of $\pi $\ and $\sigma $ bands are considered.
By virtue of unitary transformation and perturbation approach, we obtain the
analytic solution for the torsion-dependent SOI in helically coiled CNTs.
Due to the enhancement of curvature and torsion, the calculated SOI values
reach the order of meV which has been confirmed by \textit{ab initio}
electronic structure calculation.
\end{abstract}


%Subject Areas: Condensed Matter Physics, Nanophysics, Spintronics

\section{INTRODUCTION}

Carbon nanotubes (CNTs) have been the focus of intense study in the past few
decades as they exhibit remarkable properties that make them good candidates
for molecular electronic devices.$^{1}$ Lately CNTs spintronics, by
combining electronics with spintronics to inject, detect, and manipulate the
electron spin in CNTs systems, has been gradually regarded as one of the
most promising research fields.$^{2-4}$\emph{\ }Meanwhile their application
as building blocks in spintronics has been also addressed.$^{5-8}$
Nonetheless, the spin-orbit interaction (SOI) has been customarily
underestimated in CNTs,$^{9}$ due to the low atomic number of carbon $(Z=6)$%
. Only recently was it proven experimentally$^{10}$ that in the spectrum of
ultraclean straight carbon nanotubes (SCNTs), the effects of this coupling
between spin and orbital degrees of freedom are clearly visible. This
observation is in agreement with previous theoretical predictions,$^{11-13}$
which argued that SOI could be significant in SCNTs due to their curvature.
Understanding the effects of this coupling is essential for the successful
manipulation of the different degrees of freedom of these systems, thereby
affecting the transport properties of electrons, so it's a crucial issue in
condensed matter physics and spintronics. To date, the spin manipulation via
SOI has been extensively studied in SCNTs.$^{14-22}$

Motivated by this, we further explore the SOI effects in the curved CNTs. In
fact, some curved CNTs (i.e. ring closure and coil-shaped CNTs, etc) have
already been founded in experiment.$^{23,24}$ Particularly, the helically
coiled carbon nanotubes (HCCNTs) firstly predicted by Ihara, Itoh, and
Kitakami,$^{25}$\ possess nonzero curvature and torsion with two kinds of
disclinations (5-membered rings and 7-membered rings), but they are absent
in the simple SCNTs.$^{26,27}$ Because of their potential applications for
nano-electronic devices and nano-electromechanical systems, the HCCNTs have
been attracting extensive interest.$^{28-30}$ Regrettably, these researches
have just focused on the electronic properties without involving the
spintronic properties. Especially, the law of additional SOI induced by the
distortion of HCCNTs and its influence on the movement of electron,\emph{\ }%
remain poorly understood. Some other critical problems should be also
thoroughly investigated, such as the chirality how to affect the SOI? Does
the SOI revised by the torsion have any of the new physical effect? The most
interesting target raised for us is to clarify the role of SOI in such
nanostructures. However, because of its complicated configuration,\emph{\ }%
the analytical calculations for SOI becomes an extremely challenging issue.
These researches have important academic significance on the design of new
mesoscopic spin electronic device.

In this paper, we calculate the SOI in HCCNTs by employing the perturbation
theory and the tight binding approximation for nearest-neighbor hopping.
These methods have been well applied to an infinite graphene and a carbon
nanotube, etc.$^{13}$ In Ref. $13$, Guinea \textit{et al.} take two atoms of
equal height along the axis of the tube (i.e. the Armchair-SCNT) as the
model, and obtain the SOI with the angle $\theta $, in the limit when the
radius of curvature is much longer than the interatomic spacing, $a\ll R$,
is given by $\theta \approx a/R$. Here we start from a more general case as
two atoms of different height (i.e. the Chiral-SCNT), and recover the angle $%
\theta $ including the chiral angle $\psi _{c}$ without the limit condition
above, and on this basis, we can further clarify the chiral effect on the
HCCNT's SOI. To accomplish this, we introduce an unitary matrix $\mathbf{U}$%
, which has bridged the coordinates between SCNT and HCCNT. By virtue of the
$\mathbf{U}$ matrix, we derive the analytical solution $\xi _{R}$ of SOI,
and find the $\xi _{R}$ for the $\pi $ bands is first order in the atomic
carbon SOI strength $\xi _{0}$ (i.e. $\xi _{R}\propto \xi _{0}$), similar to
the \textquotedblleft Rashba-type\textquotedblright\ SOI in graphene.$%
^{13,31}$ But strikingly different from graphene and SCNT, the term $\xi
_{R} $ largely depends on both the curvature $\kappa $ and the torsion $\tau
$. Due to their enhancement, the calculated SOI values reach the order of $m$%
eV. To check the validity of $\xi _{1}$, we specially compute five different
\textit{ab initio} energies of SOI for five different HCCNTs of coiled pitch
$p_{s}$, and verify the theoretical estimated SOIs, qualitatively and
quantitatively, agree well with realistic \textit{ab initio} calculation
values.

This paper is organized as follows. In section $2$, a brief introduction is
given to the details of calculation methods. Finally, we derive the analytic
solution of SOI. In section $3$, \textit{ab initio} calculations are
described, and the results compared with analytic solutions are discussed.
In the last section, we present brief summary and conclusions.

\section{GENERAL SOLUTION}

In graphene, the carbon atoms are arranged into a hexagonal lattice
connected by strong covalent bonds of $\sigma $-orbitals derived from the $%
sp^{2}$ hybridization of the atomic orbitals. The remaining $p_{x}$ orbital
(normal to the atomic plane) has a weak overlap and forms a narrow band of $%
\pi $-orbitals states. To a first approximation, the $\pi $-electron system
can be modeled as a tight-binding Hamiltonian characterized by a single
hopping matrix element between neighboring atoms, and the energy offset of
the $p$-electron states. Considering the arbitrary atom of
pentagon--heptagon pairs in HCCNTs is still surrounded by three nearest
neighbor atoms similar to the hexagon case, we reasonably choose the
tight-binding model with a two-center Slater-Koster approximation$^{32}$ for
nearest-neighbor hopping in our calculations. The tight-binding
representation of Hamiltonian is defined by the combination of the
contributions $H_{0}$ and $H_{SO}$,
\begin{equation}
H=H_{0}+H_{SO}.
\end{equation}%
Here, the spin-independent noninteracting Hamiltonian reads $%
H_{0}=\sum\limits_{i\mu s^{\prime }}c_{i\mu s^{\prime }}^{\dag }t_{i\mu
}c_{i\mu s^{\prime }}+\sum\limits_{\left\langle i\mu s^{\prime },j\mu
^{\prime }s^{\prime }\right\rangle }c_{i\mu s^{\prime }}^{\dag }t_{i\mu
,j\mu ^{\prime }}c_{j\mu ^{\prime }s^{\prime }}$, where $<i,j>$ is a
shorthand used to denote neighboring atomic sites, $<\mu ,\mu ^{\prime }>$
refers to the $s$ and $p_{x}$, $p_{y}$, $p_{z}$ atomic orbitals on each
site, and $s^{\prime }=\uparrow ,\downarrow $ the electronic spin. The term $%
t_{i\mu }$ stands for the \textquotedblleft on-site\textquotedblright\
atomic energies of 2$s$ and 2$p$ orbitals, i.e. the site-diagonal matrix
elements $t_{is}$ and $t_{ip}$, with the latter one $t_{ip}=0$. Besides, the
Wannier representation of $H_{0}$ describes electrons hopping from $j$ atom\
to another,\emph{\ }$i$. The strength of the hopping matrix element $t_{i\mu
,j\mu ^{\prime }}$ is controlled by the effective overlap of neighboring
atoms. The tight-binding representation becomes useful when the $t_{i\mu
,j\mu ^{\prime }}$ is non-vanishing, but the orbital overlap is so weak that
only nearest-neighbor hopping effectively contributes. Here we use one
parameter $V_{pp}^{\pi }$ for the nearest-neighbor hopping between the $%
p_{x} $ orbitals of the $\pi $ band, and other parameters of $V_{pp}^{\sigma
},$ $V_{sp}^{\sigma },$ and $V_{ss}^{\sigma }$ for the rest of the
intra-atomic hoppings between the atomic orbitals $s$, $p_{x}$, $p_{y}$ of
the $\sigma $ band. The SOI arises $H_{SO}=\xi _{0}$ $\mathbf{L}\cdot
\mathbf{s}$\textbf{,}$^{33,34}$ with the intra-atomic SOI constant $\xi _{0}$%
, the total atomic angular momentum operator $\mathbf{L=r\times p}$, and the
total electronic spin operator $\mathbf{s}$.

Following the approach of Ando$^{11}$ and Guinea$^{13}$ \textit{et al.}, we
analyse the hopping between the $p_{x}$ and $p_{y}$ orbitals in the $\pi $
and $\sigma $ bands for SCNT (see Fig. $1$). Here we assume that the $p_{x}$%
\emph{\ }orbitals are oriented normal to the surface,\emph{\ }the $p_{y}$\
orbitals tangent to the surface circumference,\emph{\ }and the $p_{z}$\
orbitals parallel to the tube axes. Comparing to graphene, the curvature
modifies the hopping for the\emph{\ }$p_{x\left( y\right) }$\emph{\ }%
orbitals between the two neighboring atoms, but not changes the hopping
between $p_{z}$\ orbitals, so the effective projections\emph{\ }should be
oriented along $p_{x}$ and $p_{y}$ except of $p_{z}$. Hence, the revised $%
p_{x}$-$p_{y}$ hopping Hamiltonian is given by%
\begin{eqnarray}
H_{h} &=&\sum\limits_{s^{\prime }}\left[ V_{pp}^{\pi }\cos ^{2}\theta
+V_{pp}^{\sigma }\sin ^{2}\theta \right] c_{x1s^{\prime }}^{\dag
}c_{x0s^{\prime }} \\
&&-\sum\limits_{s^{\prime }}\left[ V_{pp}^{\pi }\sin ^{2}\theta
+V_{pp}^{\sigma }\cos ^{2}\theta \right] c_{y1s^{\prime }}^{\dag
}c_{y0s^{\prime }}  \notag \\
&&+V_{sp}^{\sigma }\sin ^{2}\theta c_{x1s^{\prime }}^{\dag }c_{s0s^{\prime
}}+\sin \theta \cos \theta \left( V_{pp}^{\pi }-V_{pp}^{\sigma }\right)
\notag \\
&&\times \left( c_{x1s^{\prime }}^{\dag }c_{y0s^{\prime }}-c_{y1s^{\prime
}}^{\dag }c_{x0s^{\prime }}\right) +H.c.,  \notag
\end{eqnarray}%
in which 0 and 1 denote the two neighboring atoms considered. Strikingly
different from the case [see Fig. $1\left( a\right) $] in Ref. $13$, we take
into account two atoms of different height [see Fig. $1\left( b\right) $],
and give the angle $\theta $ between the adjacent $p_{x}$ axises [see Fig. $%
1\left( c\right) $], with comprising the chiral angle $\psi _{c}\in \left[
0,30^{0}\right] $ for the following expression,%
\begin{equation}
\{%
\begin{array}{c}
\sin \theta =\frac{a}{\sqrt{a^{2}+\left( R/\cos \psi _{c}\right) ^{2}}}, \\
\cos \theta =\frac{\left( R/\cos \psi _{c}\right) }{\sqrt{a^{2}+\left(
R/\cos \psi _{c}\right) ^{2}}},%
\end{array}%
\end{equation}%
where the symbols of $a$ and $R$, respectively, represent the lattice
spacing and the tube radius. Eq. $\left( 3\right) $ could help us to further
discuss the chiral effects on the SOI.

\begin{center}
\centering\includegraphics[width=3.5in]{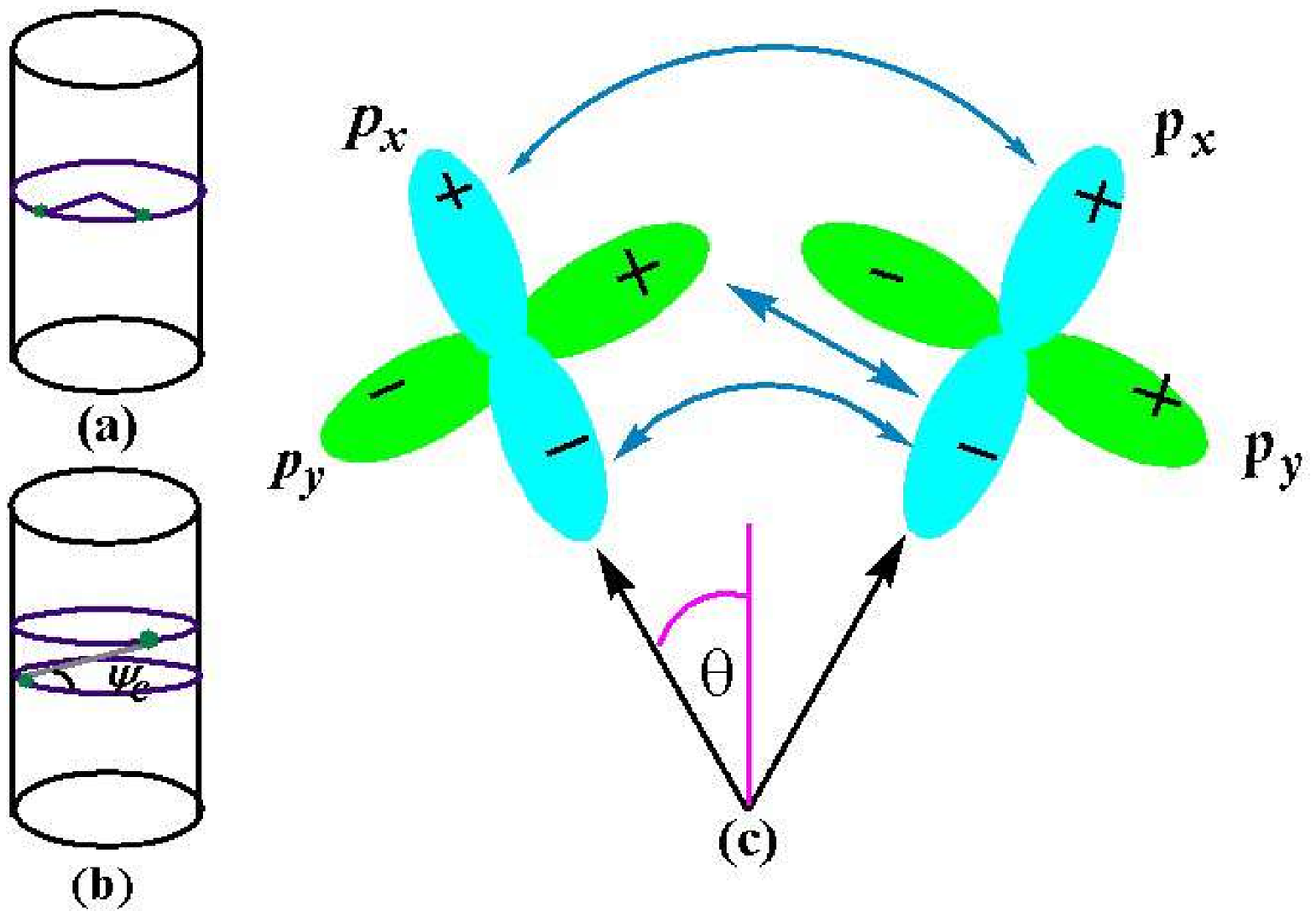}\vskip0in %\figcaption
{}
\end{center}

\begin{flushleft}
{FIG. 1. (Color online) Schematic representation of the curvature effect on
the transfer integrals between the orbitals of two nearest-neighbor atoms of
}$\left( a\right) $ equal height and $\left( b\right) $ different height {in
SCNT. }$\left( c\right) $ {The $p_{x}$ orbitals are not parallel anymore, so
that a mixing with the orbitals $p_{y}$ building the }$\sigma $ bonds takes
place{. The blue arrows stand for the different hoppings described in the
text.}
\end{flushleft}

\begin{center}
\setlength{\intextsep}{4in plus .2in minus 0.2in}
\end{center}

The Hamiltonian of Eq. $\left( 2\right) $ actually demonstrates the
transition of $p_{x}$-$p_{y}$ orbitals in SCNT with the coordinate of $%
(x,y,z)$, but not in HCCNT with $(x^{\prime },y^{\prime },z^{\prime })$.
Aiming at this problem, we make an ansatz, that there exists an unitary
matrix $\mathbf{U}$ enables us to replace the $(x,y,z)$ with the $(x^{\prime
},y^{\prime },z^{\prime })$ in Eq. $\left( 2\right) $:%
\begin{equation}
\left( c_{x},c_{y},c_{z}\right) ^{T}=\mathbf{U}\left( c_{x}^{\prime
},c_{y}^{\prime },c_{z}^{\prime }\right) ^{T},
\end{equation}%
thereby leading to an transformation of the $H_{h}$ from SCNT to HCCNT,
which is undoubtedly the crucial step in our calculations. Thereafter,
through analysing the connection between the {coordinate transformation} and
the topological structure as illustrated in Fig. $2$, we find out the matrix
$\mathbf{U}$ as

\begin{equation}
\mathbf{U}=\left(
\begin{array}{ccc}
\cos \alpha _{1}\cos \beta _{1} & -\cos \alpha _{2}\sin \beta _{2} & -\sin
\phi \cos \left( \frac{\pi }{4}+\eta \right) \\
-\cos \alpha _{1}\sin \beta _{1} & \cos \alpha _{2}\cos \beta _{2} & -\sin
\phi \sin \left( \frac{\pi }{4}+\eta \right) \\
\sin \alpha _{1} & \sin \alpha _{2} & \cos \phi%
\end{array}%
\right) ,
\end{equation}%
where the $3\times 3$ matrix $\mathbf{U}$ obeys $\mathbf{UU}^{\dag }=\mathbf{%
U}^{\dag }\mathbf{U}=\mathbf{I}$ and all parameters are denoted in Fig. $2$.
Specifically, the process of $\mathbf{U}$ transformation consists of two
operations, i.e., around the dash line and the axis $z$, sequentially
rolling and spining the local coordinate $(x,y,z)$ by $\eta $ and $\phi $
(see Fig. $2$) for the atoms previously distributing in SCNT. Ultimately, we
get the local coordinate $(x^{\prime },y^{\prime },z^{\prime })$ for the
atoms redistributing in HCCNT. Note, the axis $z^{\prime }$ is required to
be tangent to the central axis of HCCNT, and the type of HCCNT is completely
determined by the inclination angle of helix $\phi $ belonging to $\left[
0,90^{0}\right] $, with two limits of 0$^{0}$ and 90$^{0}$, respectively,
corresponding to the straight CNTs and the torus CNTs.

\begin{center}
%\begin{minipage}{0.85\textwidth}
\centering\includegraphics[width=5in]{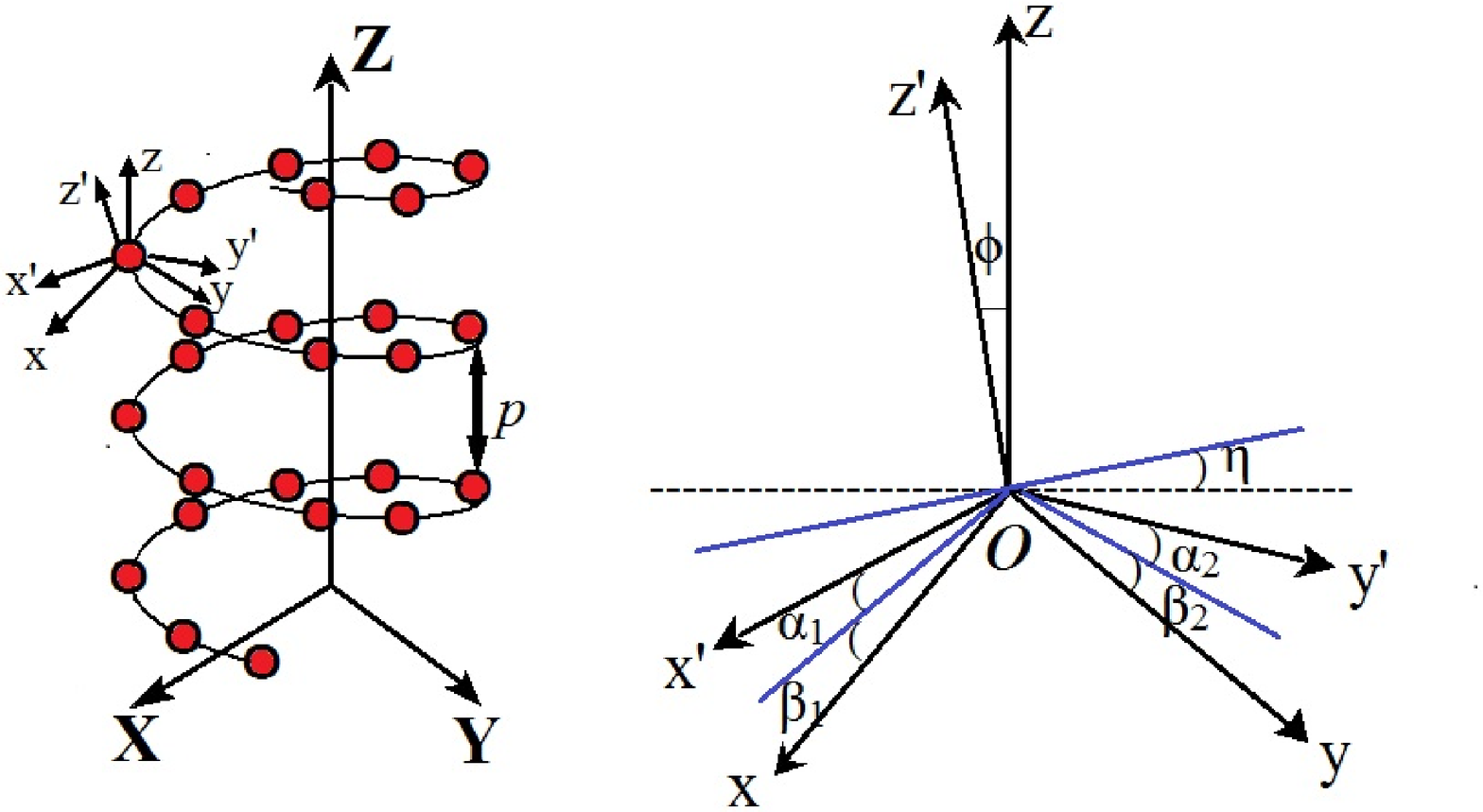}\vskip0in
%\figcaption{(Color\ online)
\end{center}

\begin{flushleft}
{FIG. 2. (Color online) Schematic view of the coordinate transformation from
the }$(x,y,z)$ of {SCNT to the }$(x^{\prime },y^{\prime },z^{\prime })$ of {%
HCCNT. The atoms of }one spiral line in HCCNT are employed to interpret the
process. The fixed coordinate of $(X,Y,Z)$, the original local coordinate of
$(x,y,z)$, and the final one of $(x^{\prime },y^{\prime },z^{\prime })$ are
established on the arbitrary atom in the spiral line.
\end{flushleft}

\begin{center}
%\end{minipage}
\setlength{\intextsep}{4in plus 0.2in minus 0.2in}
\end{center}

Substituting Eq. $\left( {4}\right) $ into Eq. $\left( {2}\right) $, and
projecting onto the bloch wave functions of the $\pi $ and $\sigma $ bands
at the $K\left( K^{\prime }\right) $ point, we obtain the Hamiltonian of SOI
for HCCNT
\begin{eqnarray}
H_{hK(K^{\prime })} &=&\frac{\sqrt{6}}{4}\cos \alpha _{1}\cos \alpha _{2}%
\left[ \left( V_{pp}^{\pi }-V_{pp}^{\sigma }\right) \sin \left( \beta
_{1}+\beta _{2}-2\theta \right) +\left( V_{pp}^{\pi }+V_{pp}^{\sigma
}\right) \sin \left( \beta _{2}-\beta _{1}\right) \right] \\
&&\times \int d^{2}\mathbf{r}\{\cos \left( \frac{\alpha _{0}}{2}\right)
[\Psi _{\pi AK(K^{\prime })\uparrow }^{\dag }\left( \mathbf{r}\right) \Psi
_{\sigma 1(2)BK(K^{\prime })\uparrow }\left( \mathbf{r}\right) +\Psi _{\pi
BK(K^{\prime })\uparrow }^{\dag }\left( \mathbf{r}\right) \Psi _{\sigma
1(2)AK(K^{\prime })\uparrow }\left( \mathbf{r}\right) ]  \notag \\
&&+\sin \frac{\alpha _{0}}{2}[\Psi _{\pi AK(K^{\prime })\uparrow }^{\dag
}\left( \mathbf{r}\right) \Psi _{\sigma 2(1)BK(K^{\prime })\uparrow }\left(
\mathbf{r}\right) +\Psi _{\pi BK(K^{\prime })\uparrow }^{\dag }\left(
\mathbf{r}\right) \Psi _{\sigma 2(1)AK(K^{\prime })\uparrow }\left( \mathbf{r%
}\right) ]\}+H.c.,  \notag
\end{eqnarray}%
in which $\Psi _{\pi \left( \sigma \right) }$ stands for the component of $%
\pi \left( \sigma \right) $ band, and the $\alpha _{0}$ is determined by
basic energy parameters (i.e. $V_{ss}^{\sigma }$, $V_{sp}^{\sigma }$, $%
V_{pp}^{\sigma }$, etc).$^{13}$ Following the unitary requiring of $\mathbf{U%
}$ matrix and the approach of differential geometry,$^{35,36}$ we build the
complicated relations to connect the angle parameters (i.e. $\alpha $ and $%
\beta $, etc) and the characteristic parameters (i.e. curvature $\kappa $,
torsion $\tau $, coiled pitch $p=2\pi h$, inclination angle of helix $\phi $%
, and coil radius $r_{0}$, etc) as follows,%
\begin{eqnarray}
\alpha _{1} &=&\arcsin \left[ \sin \left( \frac{\pi }{4}-\eta \right) \kappa
\sqrt{h^{2}+r_{0}^{2}}\right] , \\
\alpha _{2} &=&\arcsin \left[ \sin \left( \frac{\pi }{4}+\eta \right) \kappa
\sqrt{h^{2}+r_{0}^{2}}\right] ,  \notag \\
\beta _{1} &=&\frac{1}{2}\left\{ \arccos \left[ \lambda \left( 1-\frac{\cos
^{2}\phi }{mn}\right) ^{1/2}\right] +\arccos \left( \sqrt{\frac{\cos
^{2}\phi }{mn}}\right) \right\} -\eta ,  \notag \\
\beta _{2} &=&\frac{1}{2}\left\{ \arccos \left( \sqrt{\frac{\cos ^{2}\phi }{%
mn}}\right) -\arccos \left[ \lambda \left( 1-\frac{\cos ^{2}\phi }{mn}%
\right) ^{1/2}\right] \right\} +\eta ,  \notag
\end{eqnarray}%
with%
\begin{eqnarray}
\phi &=&\arctan \frac{r_{0}}{h}, \\
\eta &=&\arctan \left( \frac{\sqrt{3}a}{2r_{0}}\cos \psi _{c}\right) ,
\notag \\
\lambda &=&\frac{1+\tau ^{2}\left( h^{2}+r_{0}^{2}\right) }{1-\tau
^{2}\left( h^{2}+r_{0}^{2}\right) },  \notag \\
m &=&1-\sin ^{2}\left( \frac{\pi }{4}-\eta \right) \kappa ^{2}\left(
h^{2}+r_{0}^{2}\right) ,  \notag \\
n &=&1-\sin ^{2}\left( \frac{\pi }{4}+\eta \right) \kappa ^{2}\left(
h^{2}+r_{0}^{2}\right) .  \notag
\end{eqnarray}%
From Eqs. $\left( 7\right) $ and $\left( 8\right) $, two realistic facts are
revealed as: 1) The hopping term of Eq. $\left( {6}\right) $ largely depends
on the characteristic parameters; 2) The hopping term is caused by intrinsic
curvature and torsion. Assuming the energies of $\sigma $ bands are well
separated from the energy of the $\pi $ bands [$t_{\pi }=0$ at the $K\left(
K^{\prime }\right) $ point], we finally derive the effective Hamiltonian
acting on the states of the $\pi $ band from Eq. $\left( 6\right) $ by using
second-order perturbation theory,%
\begin{equation}
H_{\pi K\left( K^{\prime }\right) }=-i\xi _{R}\int d^{2}\mathbf{r}[\pm \Psi
_{\pi AK\left( K^{\prime }\right) \uparrow \left( \downarrow \right) }^{\dag
}\left( \mathbf{r}\right) \Psi _{\pi BK\left( K^{\prime }\right) \downarrow
\left( \uparrow \right) }\left( \mathbf{r}\right) \mp \Psi _{\pi BK\left(
K^{\prime }\right) \downarrow \left( \uparrow \right) }^{\dag }\left(
\mathbf{r}\right) \Psi _{\pi AK\left( K^{\prime }\right) \uparrow \left(
\downarrow \right) }\left( \mathbf{r}\right) ],
\end{equation}%
in which%
\begin{equation}
\xi _{R}=\frac{\xi _{0}V_{1}\cos \alpha _{1}\cos \alpha _{2}}{2\left(
2V_{1}^{2}+V_{2}^{2}\right) }\left[ \left( V_{pp}^{\pi }-V_{pp}^{\sigma
}\right) \sin \left( \beta _{1}+\beta _{2}-2\theta \right) +\left(
V_{pp}^{\pi }+V_{pp}^{\sigma }\right) \sin \left( \beta _{2}-\beta
_{1}\right) \right] ,
\end{equation}%
with $V_{1}=\left( t_{is}-t_{ip}\right) /3$ and $V_{2}=\left( V_{ss}^{\sigma
}+2\sqrt{2}V_{sp}^{\sigma }+2V_{pp}^{\sigma }\right) /3$. The
geometry-dependent term $\xi _{R}$ demonstrate that the hopping term of Eq. $%
\left( 6\right) $ causes an mixing of the $\pi $ and $\sigma $ orbitals,
thereby modifying the SOI in graphene and SCNT.{\ }This process is quite
sensitive to the deformations of helical lattice along the bond direction
between the different atoms where the $p$ part of the $sp^{2}$ orbitals.

Defining a $4\times 4$ spinor%
\begin{equation}
\Psi _{\pi K\left( K^{\prime }\right) }=\left(
\begin{array}{c}
\Psi _{A\uparrow \left( \mathbf{r}\right) } \\
\Psi _{A\downarrow \left( \mathbf{r}\right) } \\
\Psi _{B\uparrow \left( \mathbf{r}\right) } \\
\Psi _{B\downarrow \left( \mathbf{r}\right) }%
\end{array}%
\right) _{\pi K\left( K^{\prime }\right) },
\end{equation}%
we rewrite Eq. $(9)$ in the following \textquotedblleft
Rashba-type\textquotedblright\ interaction form:$^{13,31,37}$%
\begin{eqnarray}
H_{R\pi K\left( K^{\prime }\right) } &=&-i\xi _{R}\int d^{2}\mathbf{r}\Psi
_{\pi K\left( K^{\prime }\right) }^{\dag }\left( \mathbf{r}\right) \left(
\pm \mathbf{\sigma }_{+}\mathbf{s}_{\pm }\mp \mathbf{\sigma }_{-}\mathbf{s}%
_{\mp }\right) \Psi _{\pi K\left( K^{\prime }\right) }\left( \mathbf{r}%
\right) \\
&=&\frac{\xi _{R}}{2}\int d^{2}\mathbf{r}\Psi _{\pi K\left( K^{\prime
}\right) }^{\dag }\left( \mathbf{r}\right) \left( \mathbf{\sigma }_{x}%
\mathbf{s}_{y}+\mathbf{\lambda }_{z}\mathbf{\sigma }_{y}\mathbf{s}%
_{x}\right) \Psi _{\pi K\left( K^{\prime }\right) }\left( \mathbf{r}\right) .
\notag
\end{eqnarray}%
Here the $\mathbf{\sigma }_{\alpha }$ Pauli matrices act in the $A(B)$ space
with $\mathbf{\sigma }_{z}$ eigenstates localized on the $A(B)$ sublattice, $%
\mathbf{\lambda }_{z}=\pm 1$ describing states at the $K(K^{\prime })$
points, and the $\mathbf{s}_{\alpha }$ are Pauli matrices acting on the
electron's spin.

Our results clearly show that the effective SOI $\xi _{R}$ for the $\pi $
bands is first order in the atomic carbon SOI strength $\xi _{0}$, similar
to the \textquotedblleft Rashba-type\textquotedblright\ form in graphene.$%
^{13,31,37}$ For completeness, we add the quite weak intrinsic SOI$^{13}$
(without considering the effects of curvature and torsion) $\xi _{int}\simeq
\left( 3\xi ^{2}/4V_{1}\right) \left( V_{1}/V_{2}\right) ^{4}\sim 1.0$ $\mu $%
eV which has a leading contribution proportional to $\xi ^{2}$. Therefore,
the total SOI strength $\xi ^{\ast }$ should be further estimated by $\xi
^{\ast }=\xi _{R}+\xi _{int}.$ Taking five different coiled pitches $p\sim $
7.4, 7.6, 7.8, 8.1, 8.3 $%
%TCIMACRO{\U{212b} }%
%BeginExpansion
\text{\AA}
%EndExpansion
$m, we get the corresponding values of SOI $\xi ^{\ast }\sim $ 0.91, 0.93,
0.95, 0.99, 1.01 $m$eV, respectively. Strictly speaking, the geometry
factors (i.e., $a$, $r_{0}$, $R$, etc) are closely related to the coiled
pitch $p$, and that is mean, the calculations of SOI energies for different $%
p_{s}$ should match different factors. Whereas, the variation range of $p$
is so small that we can neglect the differences in the $p$-dependent factors.

Figure $3$ directly presents a positive correlation between the
tight-binding {SOI energies }$\xi ^{\ast }$ and the torsion ${\tau }$ with
the given parameters above{. }In stark contrast to the SOI of graphene (i.e.
the order of $\mu $eV), the SOI $\xi ^{\ast }$ of HCCNT has been notably
improved owing to the effects of curvature and torsion resulting from the
structure itself.

\begin{center}
\centering \includegraphics[width=5in]{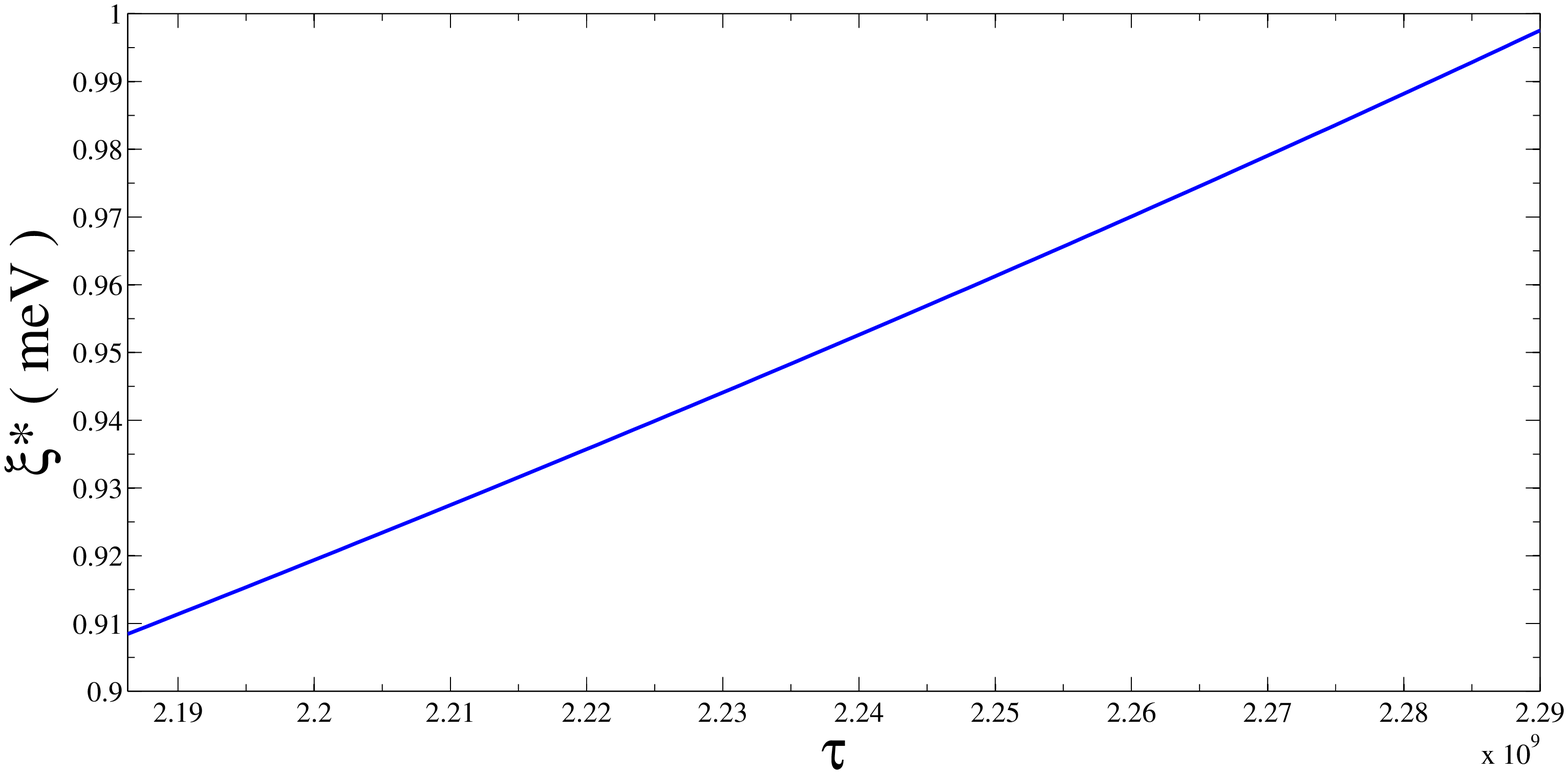}\vskip 0in %\figcaption
\end{center}

\begin{flushleft}
{FIG. 3. The tight-binding SOI energies }$\xi ^{\ast }$ {of HCCNTs\ as a
function of the torsion }${\tau }${. }The $\xi ^{\ast }$ has been estimated
using the parameters as $\xi _{0}=12$ $m$eV for the SOI constant of carbon; $%
t_{s}=-7.4$ eV, $t_{p}=0.0$ eV for the on-site energies; $V_{ss}^{\sigma
}=-3.63$ eV, $V_{sp}^{\sigma }=4.2$ eV, $V_{pp}^{\sigma }=5.38$ eV, and $%
V_{pp}^{\pi }=-2.24$ eV for the intralayer nearest-neighbor interactions
between the 2$s$, 2$p_{x}$, 2$p_{y}$, 2$p_{z}$ orbitals. Other parameters
are $V_{1}=2.47$ eV, $V_{2}=6.33$ eV, $a=1.42$ $%
%TCIMACRO{\U{212b} }%
%BeginExpansion
\text{\AA}
%EndExpansion
$m, $r_{0}=2.0$ $%
%TCIMACRO{\U{212b} }%
%BeginExpansion
\text{\AA}
%EndExpansion
$m, and $R=1.7$ $%
%TCIMACRO{\U{212b} }%
%BeginExpansion
\text{\AA}
%EndExpansion
$m.
\end{flushleft}

\begin{center}
\setlength{\intextsep}{4in plus .2in minus 0.2in}
\end{center}

\section{MODEL AND \textit{AB INITIO} CALCULATIONS}

\subsection{Model}

We construct the primitive cell of HCCNT with 120 carbon atoms $\left(
C_{120}^{HC}\right) $ by tiling the optimized pattern of torus CNT $\left(
C_{120}^{T}\right) $.\textit{\ }Along the radius of curvature,\textit{\ }the
torus is cut into small pieces,\textit{\ }which are stretched toward the
fiber axis\textit{\ }and combined continuously to obtain the initial atomic
positions of the helical structures.\textit{\ }Hence, the helix is created
so that one pitch contains one torus. Along the outer ridge line of helices C%
$_{120}^{HC}$,\ fivefold rings appear to create positive curvature in the
same fashion as in the corresponding toroidal structure.\emph{\ }Besides,
along the inner ridge line, sevenfold rings appear in representing
negatively curved surface. Figure $4$ shows an example of a helix coiled
nanotube generated by the computer. \textit{Ab initio} Molecular-dynamics
simulations gives the optimal and thermodynamically stable C$_{120}^{HC}$ .

\begin{center}
\centering \includegraphics[width=1.5in]{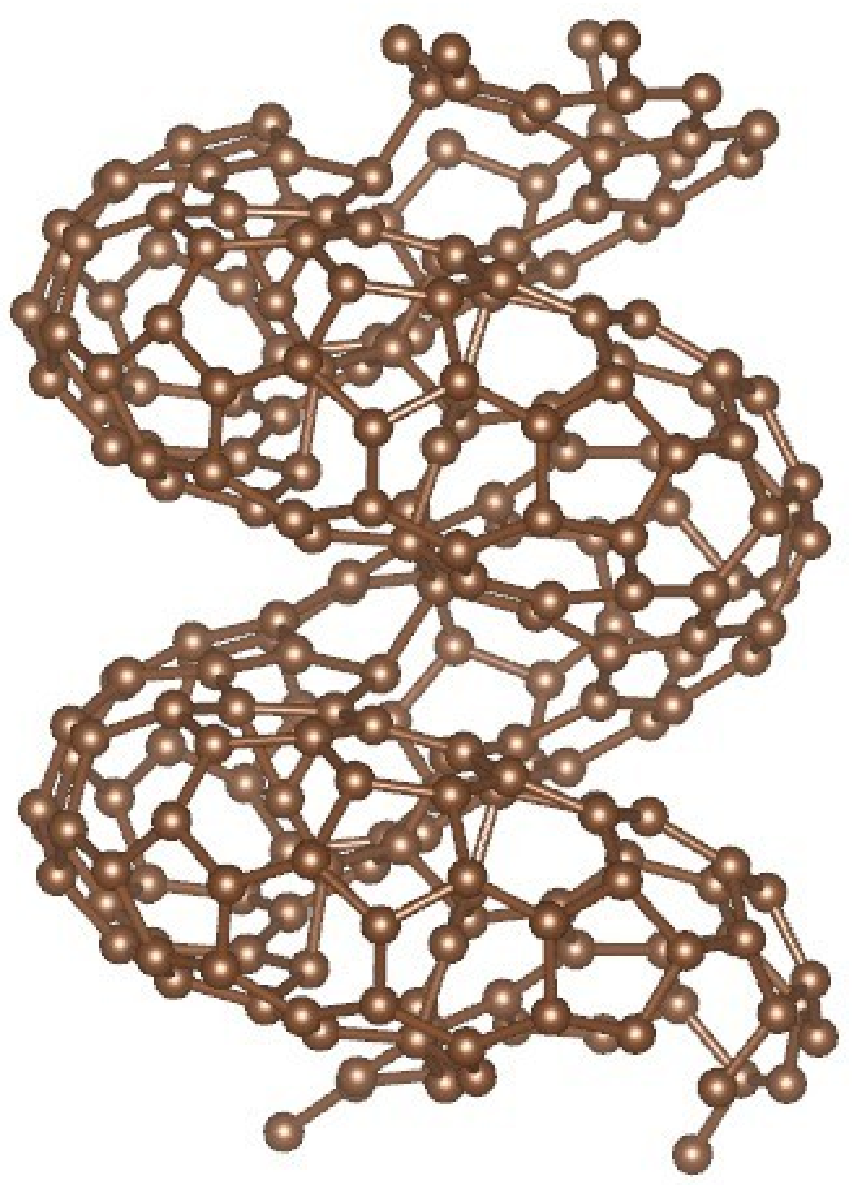}\vskip 0in %\figcaption
\end{center}

\begin{flushleft}
{FIG. 4. (Color online) The structure of a helically coiled nanotube }C$%
_{120}^{HC}${\ determined by ab initio molecular dynamics simulation (two
pitch length is shown). }
\end{flushleft}

\begin{center}
\setlength{\intextsep}{4in plus .2in minus 0.2in}
\end{center}

In present study, the helical structure of C$_{120}^{HC}$ are left handed;\
however, it is possible to form right-handed helices.\emph{\ }The optimized
ground-state structure was finally derived from \textit{ab initio}
molecular-dynamics simulations with the ultrasoft pseudopotential. The two
values of lowest cohesive energies per atom between helix C$_{120}^{HC}$ and
its corresponding toroidal structure C$_{120}^{T}$ are almost the same:
-7.37 eV, -7.39 eV, respectively.\emph{\ }This may be due to the fact that
the local and global networks of the rings of these structures are
originally similar to each other.\ Besides,\emph{\ }the cohesive energy of
the fullerene C$_{60}$\ is -7.55 eV/atom and that of the graphite sheet is
-7.44 eV/atom.\emph{\ }These facts sufficiently demonstrate that the C$%
_{120}^{HC}$ is energetically stable.\textit{\ }In particular, using the
quantum potential instead of the empirical one, we believe that the
qualitative predictions mentioned below are reasonable.

\subsection{\textit{Ab initio} calculations}

We have performed realistic \textit{ab initio} electronic structure
calculations$^{38}$ for C$_{120}^{HC}$ using the projector augmented wave
(PAW)$^{39}$ method with a Perdew--Burke--Ernzerhof (PBE) generalized
gradient approximation (GGA)$^{40}$ density functional in order to partly
test the quantitative accuracy of the conclusions reached here about SOI
based on a simplified electronic structure model. The calculations were
performed using VASP (Vienna Ab initio Simulation Package).$^{41}$ In VASP,
the SOI are implemented in the PAW method which is based on a transformation
that maps all electron wave functions to smooth pseudowave functions. All
physical properties are evaluated using pseudowave functions.

Figure $5$ compares the \textit{ab initio} and tight-binding SOI energies
with respect to the coiled pitch{\ $p$}. According to Eq. $(10)$, we find
that the strength of SOI approximately linearly increases with the $p$
increasing from 7.4 to 8.3 $%
%TCIMACRO{\U{212b} }%
%BeginExpansion
\text{\AA}
%EndExpansion
$m as shown with the solid (black) line. The dashed (red) line matches five
different \textit{ab initio}\emph{\ }SOI energies for five different HCCNTs
with $p\sim $ 7.4, 7.6, 7.8, 8.1, 8.3 $%
%TCIMACRO{\U{212b} }%
%BeginExpansion
\text{\AA}
%EndExpansion
$m. It demonstrates that the theoretical estimated SOIs, qualitatively and
quantitatively, agree well with realistic \textit{ab initio} calculation
results with the order of $m$eV. The emergence of minor deviation between
the two curves is because we neglect the tiny fluctuation of the $p$%
-dependent factors as discussed above.

\begin{center}
\centering\includegraphics[width=5in]{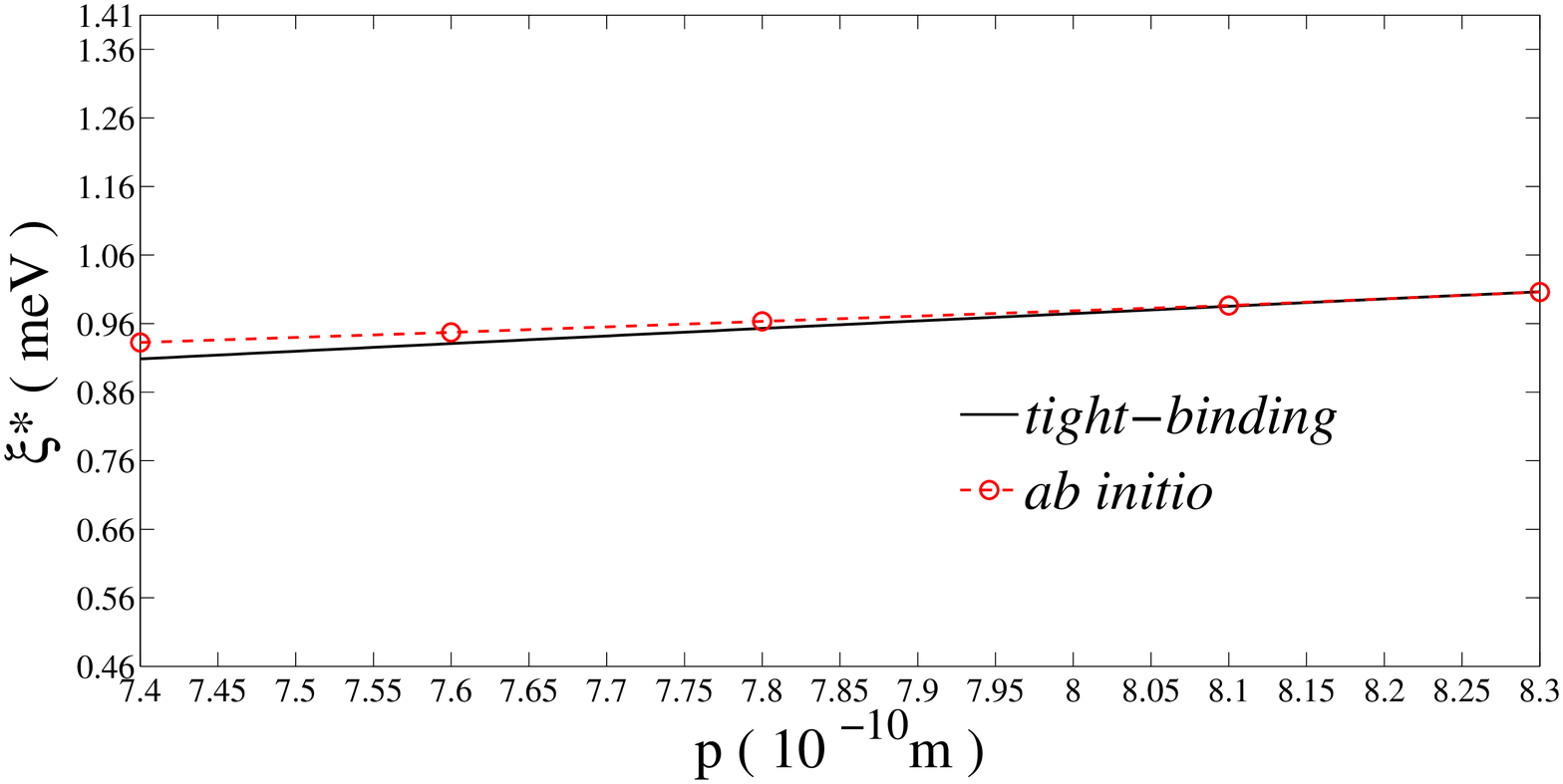}\vskip0in %\figcaption
\end{center}

\begin{flushleft}
{FIG. 5. (Color online) The total SOI energies of the helical structures }$%
\xi ^{\ast }${\ vs coil length of one pitch $p$. }The solid (black) line
stands for the tight-binding SOI energies, and the dashed (red) line for the
\textit{ab initio}\emph{\ }SOI energies.
\end{flushleft}

\begin{center}
\setlength{\intextsep}{4in plus .2in minus 0.2in}
\end{center}

\section{CONCLUSIONS}

This paper mainly reports on a theoretical study on the SOI for HCCNTs. The
SOI was treated following the approach in Refs. $11$ and $13$. The mixing of
the $\pi $ and $\sigma $ bands, due to the distortion of HCCNTs, is
automatically taken into account by this approach. We derived an analytic
solution $\xi _{R}$ for the SOI of these systems, which provides a very good
matching to the numerical calculations. Following the differential geometry
approach,$^{35,36}$ we present the relation between the $\xi _{R}$ and the
geometry factors (i.e. the curvature $\kappa $ and the torsion $\tau $).
Because of $\kappa $ and $\tau $, the value of $\xi _{R}$ reaches the order
of $m$eV, three orders of magnitude higher than that in graphene. For
completeness, we add the quite weak intrinsic SOI$^{13}$ (without
considering the effects of curvature and torsion) $\xi _{int}\simeq \left(
3\xi ^{2}/4V_{1}\right) \left( V_{1}/V_{2}\right) ^{4}\sim 1.0$ $\mu $eV to
the total SOI strength $\xi ^{\ast }$.

\section{ACKNOWLEDGMENTS}

We thank E. H. Zhang and W. X. Zhang for insightful discussions and valuable
comments on the present paper. This work is financially supported by
NSF-China under Grant Nos. 11074196, 11147117, and 11305113, and by the
Cultivation Fund of the Key Scientific and Technical Innovation Project,
Ministry of Education of China (Grant No. 708082). We also acknowledges
support from the Qualified Personnel Foundation of Taiyuan University of
Technology(QPFT)(No. tyutrc-201273a).\newline

$^*$Electronic address:zhangsl@mail.xjtu.edu.cn

$^{\dag }$Electronic address:ma.ning@stu.xjtu.edu.cn

$\bigskip $

\end{document}